\documentclass[aps,prb,twocolumn, 10pt, superscriptaddress]{revtex4-2}
\usepackage{graphicx}
\usepackage{amsmath}
\usepackage{amssymb}
\usepackage{mathtools}
\usepackage{xcolor}
\usepackage[colorlinks=true, citecolor=blue, urlcolor=blue, linkcolor=blue,bookmarks=false,hypertexnames=true]{hyperref} 
\usepackage[T1]{fontenc} 

\begin{document}
\graphicspath{}

\title{Atomistic theory of moir\'e Hofstadter's butterfly in magic-angle graphene}

\author{Alina Wania Rodrigues} \thanks{awaniaro@uottawa.ca}
\affiliation{Department of Physics, University of Ottawa, Ottawa, Ontario, K1N 6N5, Canada}

\author{Maciej Bieniek} \thanks{maciej.bieniek@pwr.edu.pl} 
\affiliation{Institut f\"ur Theoretische Physik und Astrophysik, Universit\"at W\"urzburg, 97074 W\"urzburg, Germany}
\affiliation{Institute of Theoretical Physics, Wroc\l aw University of Science and Technology, Wybrze\.ze Wyspia\'nskiego 27, 50-370 Wroc\l aw, Poland}

\author{Pawe\l\ Potasz}
\affiliation{Institute of Physics, Faculty of Physics, Astronomy and Informatics, Nicolaus Copernicus University, Grudzi{\k{a}}dzka 5, 87-100 Toru\'n, Poland}

\author{Daniel Miravet}
\affiliation{Department of Physics, University of Ottawa, Ottawa, Ontario, K1N 6N5, Canada}

\author{Ronny Thomale}
\affiliation{Institut f\"ur Theoretische Physik und Astrophysik, Universit\"at W\"urzburg, 97074 W\"urzburg, Germany}

\author{Marek Korkusi\'nski}
\affiliation{Emerging Technologies Division, National Research Council of Canada, Ottawa, ON, K1A 0R6, Canada}
\affiliation{Department of Physics, University of Ottawa, Ottawa, Ontario, K1N 6N5, Canada}

\author{Pawe\l\ Hawrylak}
\affiliation{Department of Physics, University of Ottawa, Ottawa, Ontario, K1N 6N5, Canada}

\date{\today}

\begin{abstract}

We present here a Hofstadter's butterfly spectrum for the magic angle twisted bilayer graphene obtained using  an {\it ab initio} based multi-million atom tight-binding model. We incorporate a hexagonal boron nitride substrate and out-of-plane atomic relaxation. The effects of a magnetic field are introduced via the Peierls modification of the long-range tight-binding matrix elements and the Zeeman spin splitting effects. A nanoribbon geometry is studied, and the quantum size effects for the sample widths up to 1 $\mu$m are analyzed both for a large energy window and for the flatband around the Fermi level. For sufficiently wide ribbons, where the role of the finite geometry is minimized, we obtain and plot the Hofstadter spectrum and identify the in-gap Chern numbers by counting the total number of chiral edge states crossing these gaps. Subsequently, we examine the Wannier diagrams to identify the insulating states at charge neutrality. We establish the presence of three types of electronic states: moir\'e, mixed, and conventional. These states describe both the bulk Landau levels and the edge states crossing gaps in the spectrum. The evolution of the bulk moir\'e flatband wavefunctions in the magnetic field is investigated, predicting a decay of the electronic density from the moir\'e centers as the magnetic flux increases. Furthermore, the spatial properties of the three types of edge states are studied, illustrating the evolution of their localization as a function of the nanoribbon momentum.

\end{abstract}

\maketitle

\section{Introduction}

Stacking atomically-thin materials with a relative twist or a mismatch of lattice constants leads to the system exhibiting superlattice periodicity due to the moir\'e interference between the layers. These "moir\'e materials" are intriguing physical systems where novel effects arise thanks to the high tunability of these structures, making them promising candidates for quantum simulators, especially given their capacity to access vastly different regimes of electronic matter via gate tuning \cite{SuarezMorell_Barticevic_2010, Bistritzer_MacDonald_2011a, Kim_Tutuc_2017, Cao_Jarillo-Herrero_2018a, Cao_Jarillo-Herrero_2018b, Yankowitz_Dean_2019, Lu_Efetov_2019, Uri_Zeldov_2020, Balents_Young_2020, Wong_Yazdani_2020, Zondiner_Ilani_2020, Sharpe_Goldhaber-Gordon_2019, Serlin_Young_2020, Stepanov_Efetov_2020, Wu_Andrei_2021, Das_Efetov_2021, Cao_Jarillo-Herrero_2021, Stepanov_Efetov_2021, Rozen_Ilani_2021, Po_Senthil_2018, Kang_Vafek_2018, LopesdosSantos_CastroNeto_2007, Saito_Young_2020, Bernevig_Lian_2021, Song_Bernevig_2021, Bernevig_Lian_2021b, Lian_Bernevig_2021, Bernevig_Song_2021, Kang_Vafek_2020, Jung_MacDonald_2014, Yuan_Fu_2018, Guo_Scalettar_2018, Padhi_Phillips_2018, Dodaro_Wang_2018, Liu_Yang_2018, Xu_Lee_2018, Rademaker_Mellado_2018, Koshino_Fu_2018, Thomson_Scheurer_2018, Guinea_Walet_2018, Ochi_Kuroki_2018, Isobe_Fu_2018, Xu_Balents_2018, Wu_Martin_2018, Peltonen_Heikkila_2018, Fidrysiak_Spalek_2018, Pizarro_Bascones_2019, Kang_Vafek_2019, Liu_Dai_2019, Roy_Juricic_2019, Ray_Das_2019, You_Vishwanath_2019, Tarnopolsky_Vishwanath_2019, Zhang_Senthil_2019, Pizarro_Wehling_2019, DaLiao_Xu_2019, Song_Bernevig_2019, Liu_Dai_2019a, Hu_Rossi_2019, Bultinck_Zaletel_2020, Zhang_He_2020, Xie_MacDonald_2020, Vafek_Kang_2020, Repellin_Senthil_2020, Soejima_Zaletel_2020, Wilson_Pixley_2020, Padhi_Ryu_2020, Bultinck_Zaletel_2020b, Chatterjee_Zaletel_2020, Julku_Torma_2020, Xie_Bernevig_2020, Hejazi_Balents_2021, Xie_MacDonald_2021, Xie_Regnault_2021, Potasz_MacDonald_2021, Parker_Bultinck_2021, Kumar_MacDonald_2021, Ledwith_Vishwanath_2021, Kwan_Bultinck_2021, Vahedi_TramblydeLaissardiere_2021, Lewandowski_Chowdhury_2021, Chen_Meng_2021, Kwan_Parameswaran_2021, Liu_Vishwanath_2021, Khalaf_Vishwanath_2021, Peri_Huber_2021, Chatterjee_Zaletel_2022, Liu_Liu_2022, Romanova_Vlcek_2022, Zhang_Liu_2022, Chichinadze_Chubukov_2022, Ochoa_Fernandes_2022, Kwan_Parameswaran_2022, Hofmann_Lee_2022, Song_Bernevig_2022, Schindler_Bernevig_2022, Han_Kivelson_2022, Calugaru_Bernevig_2022, Hong_Zaletel_2022, Breio_Andersen_2023, Mao_Mora_2023, Xie_Regnault_2023, Morissette_Li_2023, Pan_Meng_2023, Mandal_Fernandes_2023, Islam_Zyuzin_2023, Huang_Meng_2023, Faulstich_Lin_2023}. 

In the case of magic angle twisted bilayer graphene (MATBG), the moir\'e lattice potential leads to the formation of two flat energy bands, a valence and a conduction band, around charge neutrality. Due to the presence of spin, valley, and layer degrees of freedom in MATBG, there are multiple possibilities for ground states at partial filling of the flat bands counted in the number of electrons per moir\'e unit cell. The eight spinful flat bands are characterized using the filling fraction $\nu = [-4,4]$, where $\nu=-4$ corresponds to completely empty flat bands and $\nu=4$ to fully occupied bands. The interacting electrons occupying these bands exhibit a variety of strongly correlated phases, allowing to study the competition between interactions and topological effects. The correlated insulating states were initially observed \cite{Cao_Jarillo-Herrero_2018a} in the case of an integer number of particles per moir\'e unit cell, with superconductivity for non-integer fillings \cite{Cao_Jarillo-Herrero_2018b}. Since then, properties of MATBG have attracted an immense experimental interest \cite{Kim_Tutuc_2017, Lu_Efetov_2019, Tomarken_Ashoori_2019, Uri_Zeldov_2020, Nuckolls_Yazdani_2020, Serlin_Young_2020, Saito_Young_2020, Stepanov_Efetov_2020, Arora_Nadj-Perge_2020, Choi_Nadj-Perge_2021, Wu_Andrei_2021, Das_Efetov_2021, Liu_Li_2021, Saito_Young_2021, Xie_Yacoby_2021, Cao_Jarillo-Herrero_2021, Pierce_Yacoby_2021, Park_Jarillo-Herrero_2021, Stepanov_Efetov_2021, Lin_Li_2022, Bhowmik_Chandni_2022, Das_Efetov_2022, Polski_Nadj-Perge_2022, Yu_Feldman_2022, Tseng_Yankowitz_2022, Grover_Zeldov_2022, Diez-Merida_Efetov_2023, Tian_Bockrath_2023, Cao_Jarillo-Herrero_2018a, Cao_Jarillo-Herrero_2018b, Yankowitz_Dean_2019, Sharpe_Goldhaber-Gordon_2019, Xie_Yazdani_2019, Kerelsky_Pasupathy_2019, Jiang_Andrei_2019, Wong_Yazdani_2020, Li_He_2020, Choi_Nadj-Perge_2021b, Lu_Efetov_2021,  Finney_Goldhaber-Gordon_2022, Zhang_NadjPerge_2022, Liu_Li_2022, Hubmann_Ganichev_2022, Zondiner_Ilani_2020, Guerci_Mora_2021, Rozen_Ilani_2021, Morissette_Li_2023, Avishai_Band_2022, Lin_Li_2022}. Substantial sample-to-sample variations and conflicting results show different ground states, which are perhaps close in energy and their energetic hierarchy is influenced by subtle details. Several factors, including hexagonal boron nitride (hBN) substrate alignment \cite{Pierce_Yacoby_2021, Yu_Feldman_2022, Lin_Ni_2019, Lin_Ni_2020, Cea_Guinea_2020, Lin_Ni_2021, Cao_Jin_2021, Shi_MacDonald_2021, Mao_Senthil_2021, Shin_Jung_2021, Long_Yuan_2022, Chen_Zhang_2022, Long_Yuan_2023, Lin_Ni_2023}, strain physics \cite{Wagner_Parameswaran_2022, Padhi_Ryu_2020, Parker_Bultinck_2021} and twist angle disorder \cite{Lu_Efetov_2019, Wilson_Pixley_2020} have been suggested as potential contributors.

Current theoretical approaches to study MATBG include the continuum models \cite{Lu_Efetov_2019, Bistritzer_MacDonald_2011a, Stepanov_Efetov_2021, Xie_MacDonald_2020, Xie_MacDonald_2021, Bultinck_Zaletel_2020, Parker_Bultinck_2021, Kumar_MacDonald_2021, Bernevig_Lian_2021, Song_Bernevig_2021, Bernevig_Lian_2021b, Lian_Bernevig_2021, Bernevig_Song_2021, Xie_Regnault_2021, Repellin_Senthil_2020, Soejima_Zaletel_2020, Kang_Vafek_2020}, Hubbard-like lattice models \cite{Wong_Yazdani_2020, Zondiner_Ilani_2020, Rozen_Ilani_2021, Po_Senthil_2018, Dodaro_Wang_2018, Liu_Yang_2018, Xu_Lee_2018, Kang_Vafek_2018, Koshino_Fu_2018, Pizarro_Bascones_2019, Ochi_Kuroki_2018, Thomson_Scheurer_2018, Xu_Balents_2018, Po_Vishwanath_2019, Kang_Vafek_2019, DaLiao_Xu_2019}, and heavy fermion description \cite{Song_Bernevig_2022, Shi_Dai_2022, Singh_Vafek_2023} which, however, cannot provide an insight into the realistic edge states or atomistic defects in the sample. \textit{Ab initio} or atomistic tight-binding calculations can tackle such issues, but they prove to be challenging due to the intrinsically large scale of the problem, i.e. large number of atoms in a moir\'e unit cell for small twist angles. Some analysis of the edge states has been performed for the twisted bilayer graphene (TBG) at larger twist angles \cite{Hasegawa_Kohmoto_2013}, and with an assumption of an enlarged interlayer hopping in order to simulate the behaviour of MATBG \cite{Fujimoto_Koshino_2021, Andrade_Guinea_2023}. 

Recently, a considerable effort has been put into studying the electronic properties of MATBG in a magnetic field. The intriguing aspect lies here in the comparable MATBG moir\'e and magnetic length scales for experimentally accessible magnetic fields. Focusing on the insulating states, a rich phase diagram in the magnetic flux versus the filling factor has been observed \cite{Kim_Tutuc_2017, Lu_Efetov_2019, Tomarken_Ashoori_2019, Uri_Zeldov_2020, Nuckolls_Yazdani_2020, Serlin_Young_2020, Saito_Young_2020, Stepanov_Efetov_2020, Arora_Nadj-Perge_2020, Choi_Nadj-Perge_2021, Wu_Andrei_2021, Das_Efetov_2021, Liu_Li_2021, Saito_Young_2021, Xie_Yacoby_2021, Cao_Jarillo-Herrero_2021, Pierce_Yacoby_2021, Park_Jarillo-Herrero_2021, Stepanov_Efetov_2021, Lin_Li_2022, Bhowmik_Chandni_2022, Das_Efetov_2022, Polski_Nadj-Perge_2022, Yu_Feldman_2022, Tseng_Yankowitz_2022, Grover_Zeldov_2022, Diez-Merida_Efetov_2023, Tian_Bockrath_2023} in magnetotransport and local electronic compressibility measurements. Streda formula can be used to determine their topological properties \cite{Streda_1982, Park_Jarillo-Herrero_2021}. However, if the observed insulating states are topologically trivial (have zero Chern number), it might happen that they originate, e.g., from a combination of two states with opposite Chern numbers. This can especially occur if there is no energy gap between the flat conduction and the flat valence bands. This energy gap can be opened by breaking the inversion and time reversal symmetry at a single particle level, which in practice is done by aligning MATBG sample with hBN \cite{Pierce_Yacoby_2021, Serlin_Young_2020, Xie_Yacoby_2021, Park_Jarillo-Herrero_2021}. 

For magnetic fields around 27 Tesla entire flux quantum per moir\'e unit cell is already achieved \cite{Bistritzer_MacDonald_2011} and physics related to the Hofstadter's fractal structure of the electronic gaps can be observed. Numerous aspects have already been elucidated through the effective models \cite{Bistritzer_MacDonald_2011, Choi_Kim_2011, Zhang_Senthil_2019b, Hejazi_Balents_2019, Lian_Bernevig_2020, Herzog-Arbeitman_Bernevig_2020, Andrews_Soluyanov_2020, Sheffer_Stern_2021, Lian_Bernevig_2021, Lian_Bernevig_2021b, Kang_Vafek_2021, Hwang_Yang_2021, Benlakhouy_Vogl_2022, Guan_Yazyev_2022, Do_Chang_2022, Herzog-Arbeitman_Bernevig_2022a, Herzog-Arbeitman_Bernevig_2022b, Paul_Fu_2022, Wang_Vafek_2022, Wagner_Parameswaran_2022, Navarro-Labastida_Naumis_2023, Singh_Vafek_2023, Herzog-Arbeitman_Bernevig_2022c, Shaffer_Santos_2022}.  Initial studies concentrated on the general accessibility of the Hofstadter physics with experimentally available magnetic fields \cite{Bistritzer_MacDonald_2011, Moon_Koshino_2012, Hasegawa_Kohmoto_2013}. The topological aspects of the flat band Landau levels (LLs) have been analyzed \cite{ Song_Bernevig_2019, Hejazi_Balents_2019, Lian_Bernevig_2020, Herzog-Arbeitman_Bernevig_2020, Kang_Vafek_2020, Fujimoto_Koshino_2021, Lian_Bernevig_2021, Hwang_Yang_2021, Guan_Yazyev_2022, Herzog-Arbeitman_Bernevig_2022c} and the Hofstadter butterfly in the chiral limit has been studied in Refs. \cite{Sheffer_Stern_2021, Benlakhouy_Vogl_2022}. The physics of flat bands in strong magnetic field and possible re-entrant superconductivity have been addressed \cite{Herzog-Arbeitman_Bernevig_2022a, Herzog-Arbeitman_Bernevig_2022b, Herzog-Arbeitman_Bernevig_2022c, Guan_Kruchkov_2022, Shaffer_Santos_2022}. 
The atomistic tight-binding calculations of TBG were able to confirm the effective model results in the large magnetic field regime as well as provide insight into the edge states. These calculations mainly focus on larger twist angles, which result in a smaller moir\'e unit cell \cite{Moon_Koshino_2012, Wang_Chou_2012, Moon_Koshino_2013, Guan_Kruchkov_2022}. In order to investigate the electronic properties of MATBG at the atomistic scale, a large sample is needed ($\sim 1 \mu $m), simulating which involves costly computation.

Our multi-million atom simulation sheds light on the MATBG's on top of hBN in a magnetic field behaviour within experimentally feasible scenarios. We consider realistic sample sizes consisting of up to one million atoms, allowing us to relate the obtained Hofstadter spectrum to the experimental observations and gain insight into the microscopic properties of the wave functions for both bulk and edge states. This establishes a solid foundation for future investigations of the magnetic phase diagram, incorporating both magnetic fields and electron-electron interactions. The paper is organized as follows. In \hyperref[s2]{Section II} we introduce an atomistic model of twisted bilayer graphene, define a unit cell for commensurate twist angles and nanoribbon geometries, and describe an  \textit{ab initio}-based tight-binding model. Then we proceed to present our theoretical framework for investigating the electronic properties of MATBG under the influence of a magnetic field. \hyperref[s3]{Section III} is devoted to the study of the electronic structure of MATBG. In the remaining sections we focus on MATBG in the nanoribbon geometry and begin our analysis in \hyperref[s4]{Section IV} studying the size effects in the electronic structure. We subsequently introduce the effect of the magnetic field in \hyperref[s5]{Section V}, where we present the results in the form of a Hofstadter spectrum. This spectrum forms the basis for the construction of  Wannier diagrams, as discussed in \hyperref[s6]{Section VI}. Turning our focus to microscopic properties, \hyperref[s7]{Section VII} examines the bulk and edge wave functions. Here, we identify three distinct types of states, namely the conventional, moiré, and mixed states. Concluding our analysis, \hyperref[s8]{Section VIII} provides an analysis of bulk and edge states, studying their evolution as a function of the magnetic field and nanoribbon 1D momentum.
\section{Model and theoretical methods}
\label{s2}
\subsection{Structural properties of twisted bilayer graphene} 
\begin{figure}\
\includegraphics[scale=0.18]{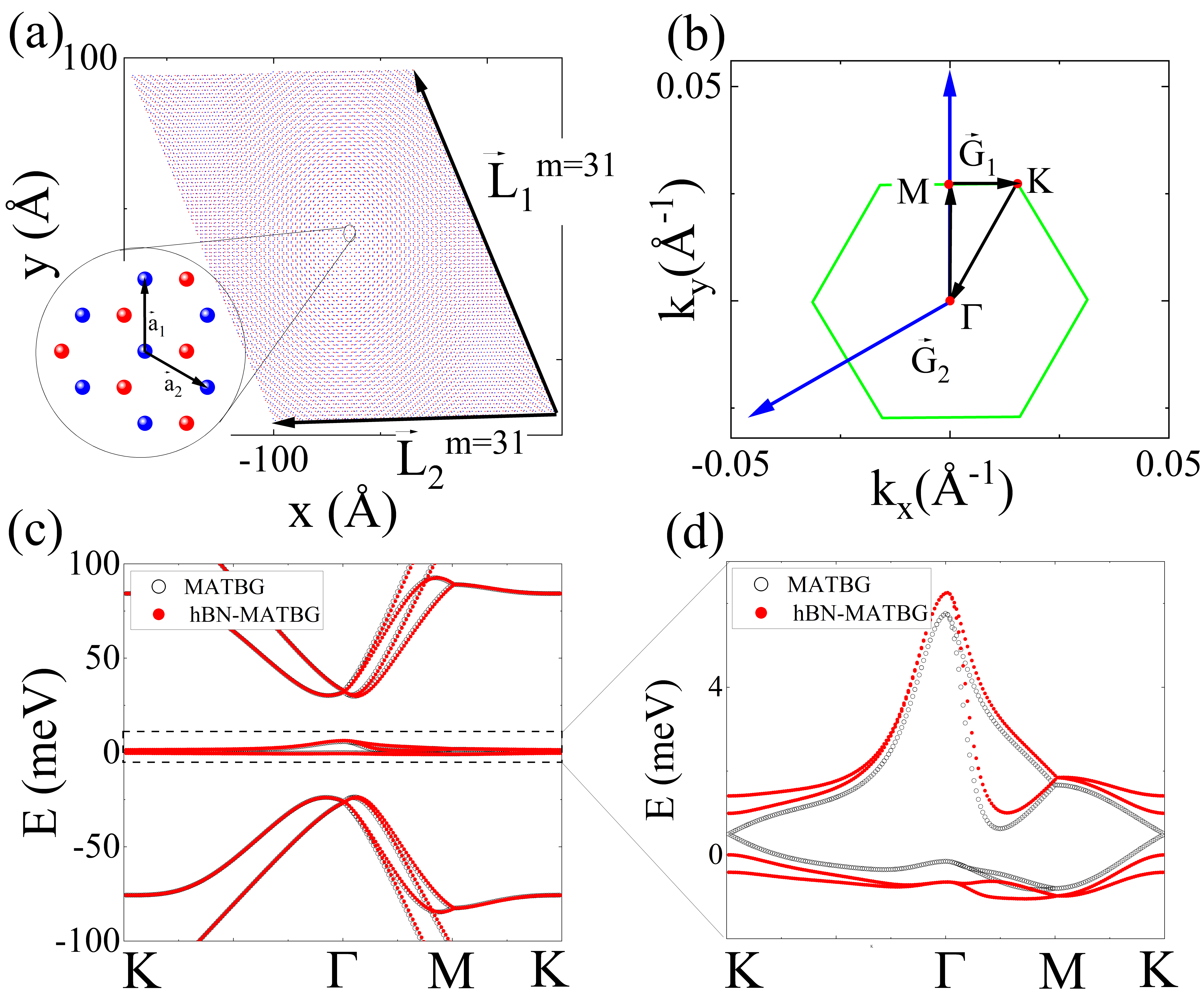}\
\caption{Structural and electronic properties of moir\'e lattice. (a) Moir\'e unit cell for MATBG consisting of 11908 atoms, spanned by the superlattice vectors $\vec{L}_1$ and  $\vec{L}_2$. The inset provides a zoom-in on a graphene structure with A and B atoms denoted by blue and red dots. The primitive lattice vectors $\vec{a}_1$ and $\vec{a}_2$ are also indicated. (b) Moir\'e BZ defined by the reciprocal lattice vectors $\vec{G}_1$ and $\vec{G}_2$. (c) MATBG band structure within a 200 meV energy window along the $K-\Gamma-M-K$ path marked in (b). MATBG is represented by black circles, while  the dispersion for MATBG on an hBN substrate is illustrated by red dots. (d) A corresponding zoom-in on the flat band.}\
\label{fig1}
\end{figure}\
Graphene is composed of carbon atoms arranged on a honeycomb lattice, with two atoms - A and B, inside a unit cell. The primitive lattice vectors are defined as $\vec{a}_1 = (0, \sqrt{3}a)$ and $\vec{a}_2=(3a/2, -\sqrt{3}a/2)$, where $a=1.412$ \AA \ represents the carbon-carbon distance, see Fig. \ref{fig1} (a). TBG is defined as two graphene sheets in a Bernal stacking, rotated by an angle $\theta$ around the (0,0) point. If $\theta$ belongs to a set of commensurate twist angles, the structure exhibiting moir\'e periodicity and
linearly independent moir\'e primitive vectors can be defined as follows:
\begin{equation}
\begin{split}
 \vec{L}_1^{(m)} &=m\vec{a}_1 - (m+1)\vec{a}_2, \\
 \vec{L}_2^{(m)} &=- (m+1)\vec{a}_1 - (2m+1)\vec{a}_2.
\end{split}
\end{equation}
$m$ is an integer, which can be used to define $\theta$ following the formula \cite{LopesdosSantos_CastroNeto_2007} $\cos(\theta_m) = (3m^2+3m+1/2)/(3m^2+3m+1)$.

 We focus on the magic angle $\theta = 1.05^{\circ}$, which corresponds to $m=31$. The number of atoms within a moir\'e unit cell can be calculated through the relation $N_{\rm at}=4(m^2+(m+1)^2+m(m+1))$, which, for the magic angle, yields $N_{\rm at}=11 908$ atoms.

The distance between the two graphene layers depends on the relative stacking and is nonuniform in the moir\'e unit cell. We account for the lattice relaxation effects along the z-axis by varying the inter-layer distance based on the stacking configuration. The layers are closest to each other in the AB stacked regions where $z_{\rm AB} = z_{\rm min} = 3.34$ \AA. The largest distance arises in AA stackings, for which we adopt $z_{\rm AA} = z_{\rm max} = 3.61$ \AA  \cite{Uchida_Oshiyama_2014}. To determine the relaxed interlayer distance $z_{\rm relax}$, we introduce a stacking parameter $\delta z$ assigned to each atom. This parameter depends on the distance in the xy plane between the considered atom and its nearest neighbour of a different type in the other layer. Specifically, it is defined as $\delta z =a^{-1} \sqrt{(x_1-x_2)^2 + (y_1-y_2)^2}$, where $x_1$ and $y_1$ describe the position of the considered atom, and $x_2$ and $y_2$ refer to its nearest neighbour in the other layer. The factor $a^{-1}$ is  included, so that $\delta z = 0$ in  the AA stacking and $\delta z = 1$ for the AB stacking. The relaxed inter-layer distance is then calculated as $z_{\rm relaxed} = z_{\rm max} - \delta z (z_{\rm max}-z_{\rm min})$.

Using the superlattice vectors $\vec{L}_1$ and $\vec{L}_2$, we determine the corresponding reciprocal space primitive vectors $\vec{G}_1$ and $\vec{G}_2$, which are used to construct a moir\'e Brillouin zone (BZ). Using the definition $\vec{G}_i \cdot \vec{L}_j = 2\pi \cdot \delta_{ij}$ we derive $\vec{G}_1$ and $\vec{G}_2$ in the form:
\begin{equation}
\begin{split}
    \vec{G}_{1} &= \left[-\frac{2\pi}{3a}\frac{1}{3m^2+3m+1}, \frac{2\pi}{3a}\frac{\sqrt{3}(2m+1)}{3m^2+3m+1} \right], \\
    \vec{G}_{2} &= \left[-\frac{2\pi}{3a}\frac{3m+1}{3m^2+3m+1}, \frac{2\pi}{3a}\frac{\sqrt{3}(m+1)}{3m^2+3m+1} \right].
\end{split}
\end{equation}
 After obtaining $\vec{G}_1$ and $\vec{G}_2$, we proceed to construct a moir\'e BZ, in which high symmetry points $\Gamma, K, M$ can be identified, as illustrated in Fig. \ref{fig1} (b).

In order to study the effects of a magnetic field we construct a nanoribbon structure by stacking moir\'e unit cells in the $\vec{L}_{1}$-direction and applying periodic boundary conditions in the $\vec{L}_{2}$-direction.  We analyze nanoribbons composed of up to 85 moir\'e unit cells, which means that we simulate MATBG system widths of up to $10^{6}$ atoms in the direction perpendicular to the periodicity. For consistent detection of the edge and bulk states, we define the edge as encompassing 10\% of the ribbon's width, at both the lower and upper ends of the sample. The remaining 80\% of atoms constitute the bulk region of the structure. In order to manage the computational effort of diagonalizing matrices of such dimensions, we have used high-performance FEAST algorithm \cite{Polizzi_2009}, which offers enhanced efficiency in the computation of the eigenvalues within a designated energy interval.

The choice of a moir\'e unit cell for an infinite system is, in general, arbitrary with respect to the real space coordinate system. However, due to our finite geometry, each particular definition of the unit cell has consequences for the electronic properties of our system. To avoid cutting through the AA region, where the flat band LDOS is expected to localize, we have chosen a unit cell that centers around the AA stacked atoms and maximizes the region of the 'moir\'e center'. This is achieved by shifting the unit cell by $(\vec{L}_1 + \vec{L}_2)/6$ with respect to the AB stacked atoms at the coordinates $(0,0)$. By doing so, we have created an irregular edge, which affects the properties of our system on the order of $0.1$ meV. This effect is visible, for example, in the flat band of large nanoribbons where the states shift with respect to the band edges deduced from the infinite system.

\subsection{Tight-binding model in magnetic field}

We employ an \textit{ab initio}-based tight-binding model for $p_z$ atomic orbitals \cite{TramblydeLaissardiere_Magaud_2010, Kerelsky_Pasupathy_2019} and tunneling over all the atoms in the sample, with a Hamiltonian given by: 
\begin{equation}
    \hat{H}_{\rm TB} = \sum_{i,j} ^N \sum _\sigma t(\vec{r}_{i},\vec{r}_{j})(c_{i,\sigma}^{\dagger}c_{j,\sigma} + c_{j, \sigma}^{\dagger}c_{i, \sigma}).
\end{equation}
$N = N_{\rm UC} \cdot N_{\rm at}$ represents the total number of atoms equal to the number of unit cells multiplied by the number of atoms in the unit cell, while $c_{i, \sigma}^{\dagger}$ ($c_{i, \sigma}$) is the creation (annihilation) operator on the i-th site located at the coordinate $\vec{r}_{i}=(x_{i}, y_{i}, z_{i})$ with spin $\sigma=\{\left \uparrow, \downarrow  \right\}$. The out-of-plane coordinate $z_{i}$ is given by its relaxed value $z_{\rm relaxed}$. The hopping term is defined as \cite{TramblydeLaissardiere_Magaud_2010}:
\begin{equation}
\begin{split}
    t(\vec{r}_{i},\vec{r}_{j}) = &(1-n^2)\gamma_0 \exp \left( \lambda_1 \left( 1 - \frac{|\vec{r}_i-\vec{r}_j|}{a} \right)\right)+ \\ 
    &n^2 \gamma_1 \exp \left(\lambda_2 \left(1-\frac{|\vec{r}_i-\vec{r}_j|}{c} \right) \right),
\end{split}
\label{hopping}
\end{equation}
where $a=1.412$ \AA \  is again the carbon-carbon distance, and $c=3.36$  \AA \  is a non-relaxed interlayer distance. The direction cosine along the z-axis is denoted by $n$. Intra-layer hopping is parametrized by $\gamma_0=-2.835$ eV, while inter-layer by $\gamma_1 = 0.48$ eV. The dimensionless decay constants are $\lambda_1=3.15$  and $\lambda_2=7.50$. For our calculations, we consider interactions of atoms within a distance $\leq6a$ from the considered atom. Including additional neighbours does not cause significant quantitative changes in our results. To account for the presence of an hBN substrate, we introduce a staggered potential $\Delta=10$ meV to the bottom layer of our system. To include this in our Hamiltonian we add a diagonal part $H_{\rm hBN}=\sum_{i,\sigma}  \Delta_{i}c_{i,\sigma}^{\dagger}c_{i,\sigma}$, where $\Delta_{i} = 5$ meV if $i$ belongs to the set of A atoms, and $\Delta_{i} = -5$ meV if $i$ belongs to the set of B atoms. If $i$ belongs to the set of atoms from the top layer $\Delta_{i} = 0$ meV.

As we have already mentioned each moir\'e unit cell consists of $N_{at}$ number of atoms. Each atom repeats in the unit cells creating $N_{\rm at}$ simple Bravais sublattices. For a given wave vector $\vec{k}$ for each sublattice $n$ we can define a wavefunction:
\begin{equation}
\Psi^n_{\vec{k}}=\frac{1}{\sqrt{N_{UC}}} \sum ^{N_{UC}}_\alpha e^{i\vec{k} \cdot (\vec{\tau}_n + \vec{R}_\alpha)} \phi_z(\vec{r} - \vec{\tau}_n - \vec{R}_\alpha)
\end{equation}
Here, $\vec{\tau}_n$ refers to the position of an atom within a unit cell, while $\vec{R}_\alpha$ denotes the position of different unit cells. To construct our Hamiltonian in the second quantization we expand the field operators in sublattice wavefunctions for each wave vector $\vec{k}$, e.g., $\hat{\Psi}_{\vec{k}}^\dagger = \sum_{n}^{N_{at}} \sum_\sigma c_{{\vec{k}},n, \sigma}^\dagger \Psi_{\vec{k}}^{*n}$. Here $c_{{\vec{k}},n,\sigma}^\dagger$ ($c_{{\vec{k}},n,\sigma}$) are the creation (annihilation) operators that create (annihilate) particles with momentum $\vec{k}$ on sublattice $n$ and spin $\sigma$.
We can now define an $\hat{H}_k$ operator for each considered $\vec{k}$ value:
\begin{equation}
\hat{H}_{\vec{k}} = \int dr \hat{\Psi}^\dagger_{\vec{k}} H \hat{\Psi}_{\vec{k}} = \sum_{n,m=1}^{N_{at}} \sum_{\sigma} c^\dagger_{{\vec{k}},n, \sigma} c_{{\vec{k}},m, \sigma} \int dr \Psi_{\vec{k}}^{*n} H \Psi_{\vec{k}}^m
\end{equation}

We introduce $H^{nm}_{\vec{k}} = \int dr \Psi_{\vec{k}}^{*n} H \Psi_{\vec{k}}^m$ as a matrix in the space of sublattices which has the form:

\begin{equation}
\begin{split}
H^{nm}_{\vec{k}} &= \int dr \left[ \frac{1}{\sqrt{N_{UC}}} \sum_\alpha^{N_{UC}} e^{-i\vec{k}\cdot(\vec{\tau}_n+\vec{R}_\alpha)} \phi_z^*(\vec{r}-\vec{\tau}_n - \vec{R}_\alpha) \right]\\
&\quad \times H \left[ \frac{1}{\sqrt{N_{UC}}} \sum_\beta^{N_{UC}} e^{-i\vec{k}\cdot(\vec{\tau}_m+\vec{R}_\beta)} \phi_z(\vec{r}-\vec{\tau}_m - \vec{R}_\beta) \right]
\end{split}
\end{equation}
Now we center our system at $\vec{r}' = \vec{r}-\vec{\tau}_n - \vec{R}_\alpha$ and define $\vec{R}_\gamma = \vec{R}_\alpha - \vec{R}_\beta $. These operations lead us to a simplified expression:

\begin{equation}
\begin{split}
H^{nm}_{\vec{k}} &= \frac{1}{N_{\rm UC}}\sum_\alpha^{N_{\rm UC}}\sum_{\gamma}^{N_{\rm UC}} e^{-i\vec{k}\cdot(\vec{\tau}_n-\vec{\tau}_m)} e^{{-i\vec{k}\vec{R}_\gamma}}\\
&\quad \times \int dr \phi_z^*(\vec{r}) H \phi_z(\vec{r}+\vec{\tau}_n-\vec{\tau}_m+\vec{R}_\gamma)
\end{split}
\end{equation}
Defining $t(\vec{\tau}_n, \vec{\tau}_m+\vec{R}_\gamma) = \int dr \phi_z^*(\vec{r}) H \phi_z(\vec{r}+\vec{\tau}_n-\vec{\tau}_m+\vec{R}_\gamma) $ we can write our final form:
\begin{equation}
H^{nm}_{\vec{k}} = \sum_{\gamma}^{N_{UC}} e^{-i\vec{k}\cdot(\vec{\tau}_n-\vec{\tau}_m)} e^{{-i\vec{k}\vec{R}_\gamma}} t(\vec{\tau}_n, \vec{\tau}_m+\vec{R}_\gamma).
\end{equation}
We note that because the hopping vanishes quickly with the distance, the sum over $\gamma$ reduces to a single non-zero element from either the considered unit cell or its nearest neighbours.  Hopping terms $t(\vec{\tau}_n, \vec{\tau}_m+\vec{R}_\gamma)$ are defined in eq. \ref{hopping} and depend on the sublattice and unit cell position of the considered atoms.  $\vec{k} = (k_x, k_y)$ represents the wave vector within the 1st moir\'e BZ.

In a nanoribbon structure, we incorporate the periodicity solely in the x-direction, modifying our Hamiltonian to have the form:
\begin{equation}
\hat{H}_{{k}_{\rm 1D}} = \sum_{n,m}^{N_{\rm at}}\sum_{\sigma}  c_{k_{\rm 1D},n,\sigma}^\dagger c_{k_{\rm 1D}, m, \sigma} H^{nm}_{k_{\rm 1D}}
\end{equation}
where $k_{\rm 1D} \in (-\pi/|\vec{L}_{2}|, \pi/|\vec{L}_{2}|)$ denotes the wave vector in the one dimensional BZ of the ribbon. 

Thanks to the nanoribbon geometry perpendicular magnetic field can be introduced using Peierls substitution in the Landau gauge $\vec{B}=\vec{\nabla} \times \vec{A}$, with vector potential $\vec{A}=(-B_{\rm z}y,0,0)$:
\begin{equation}
\begin{split}
    \tilde{t}(\vec{r}_{i},\vec{r}_{j})
    = t(\vec{r}_{i},\vec{r}_{j})^{B_z=0}\exp{\left(-i\pi\frac{2eB_{\rm z}}{hc}\frac{(x_j-x_i)}{2}(y_i+y_j)\right)}.
\end{split}
\end{equation}
Here $hc/2e$ is the magnetic flux quantum $\varphi_0$. The Zeeman splitting is included by adding a diagonal part to our Hamiltonian in the form: $H_{\rm Z} = \sum_{i} \frac{1}{2}g \mu_BB_{\rm z} \left(c_{i,\uparrow}^{\dagger}c_{i,\uparrow}-c_{i,\downarrow}^{\dagger}c_{i,\downarrow}\right)$, where $\mu_B$ is the Bohr magneton and $g$ is the Land\'e factor.
In order to use the Landau gauge we perform a global rotation of  the coordinates system, ensuring strict x-axis periodicity. Due to such periodicity, we are able to include magnetic field in our nanoribbons without the restriction to specific magnetic flux values.

We notice the existence of the topologically trivial and nontrivial edge states characterised by Chern numbers. The Chern numbers are determined by counting the number of chiral edge states crossing the energy gap. The microscopic properties of the wave functions of the flat band and the edge states are characterized by the local density of states (LDOS) for a representative choice of the magnetic field. We use the following definition of LDOS:
\begin{equation}
n(x_i, y_i, E) = N_{k_{\rm 1D}}^{-1}\sum_{\lambda, k_{\rm 1D}} |\psi_{k_{1D}, \lambda}(x_i, y_i)|^2 \delta (E-E_{k_{1D},\lambda}).
\end{equation}
Here, $n$ is the LDOS for given coordinates $x_i$ and $y_i$ in a chosen energy window. $N_{\rm k_{\rm 1D}}$ is the normalization constant and relates to the number of states considered, $\lambda$ is the eigenvalue index, which together with the wave vector $k_{\rm 1D}$ enumerates the eigenvalue $E_{k_{\rm 1D},\lambda}$ and the eigenvector $\psi_{k_{1D}, \lambda}(x_i, y_i)$.

\subsection{Wannier diagrams}
To establish a connection between the Hofstadter spectrum and the experimental observations, Wannier diagrams can be used \cite{Wannier_1978}. These diagrams can be constructed by plotting the integrated charge carrier density, up to a given energy level,  versus the magnetic field $B$. This representation highlights the linear trends in energy gaps,  
offering a useful representation of electronic gaps in the Hofstadter-like spectrum, encompassing both single-particle and many-body interaction-induced Mott-like gaps. Moreover, linear fitting of the data in the Wannier diagrams aids in identifying “unconventional” states that defy a simple LLs picture and related single-particle Chern insulators. Such unconventional ground state sequences have been recently observed \cite{Pierce_Yacoby_2021, Yu_Feldman_2022} and explained by the sequential flavor filling mechanism. 
Wannier plots obtained by mapping these gaps as functions of filling and magnetic field provide a direct comparison with the compressibility and transport experiments \cite{Tomarken_Ashoori_2019, Saito_Young_2021, Pierce_Yacoby_2021, Park_Jarillo-Herrero_2021, Wu_Andrei_2021, Stepanov_Efetov_2021, Xie_Yacoby_2021, Yu_Feldman_2022} and can serve as a tool for identifying (non) interacting insulating states.

\section{Electronic structure of twisted bilayer graphene}
\label{s3}

\begin{figure*}\
\includegraphics[scale=0.6]{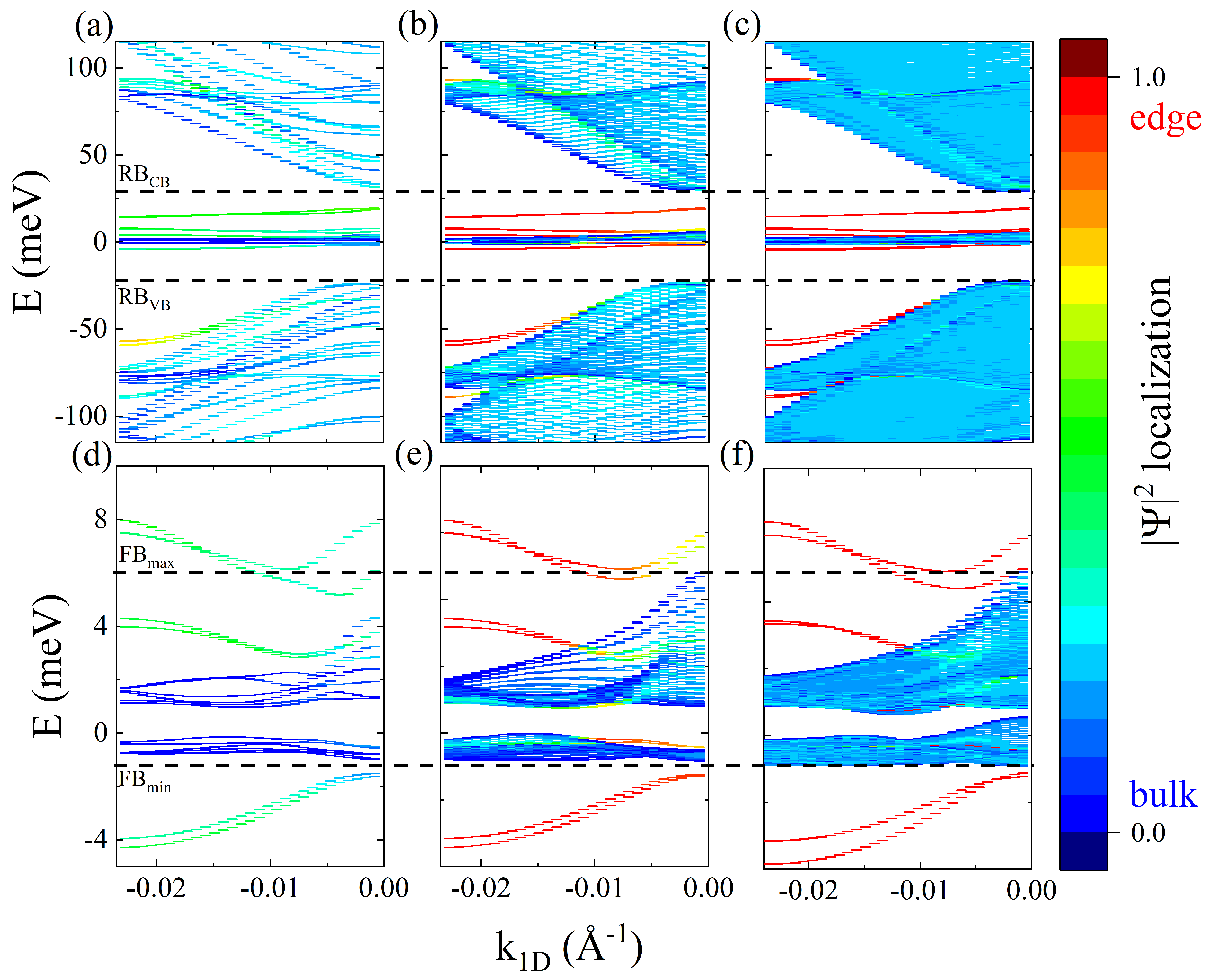}\
\caption{The electronic structure of the MATBG ribbon on an hBN substrate revealing quantization effects for the flat and remote bands. The figures (a-c) in the top panels show bands within a 200 meV energy window for different ribbon widths: (a) 5, (b) 20 and (c) 85 moir\'e unit cells. (d-f) The lower panels provide a zoom into the flat band for analogical nanoribbon widths. Color coding signifies the localization of states in real space. The dashed lines represent band edges that are deduced from the infinite system calculations. $RB_{\rm CB}$ ($RB_{\rm VB}$) refer to the top (bottom) of the remote conduction (valence) band, while $FB_{max}$ ($FB_{min}$) mark the top (bottom) of the flat band.
}\
\label{fig2}
\end{figure*}\

In Fig. 1 (c-d), we present the band structure of MATBG along the K-$\Gamma$-M-K line. In Fig. 1(c) we display the energy window that captures the flat band and the remote valence and conduction bands. In our model, we find  the flat band that is $7$ meV wide and separated by a $25$ meV gap from the remote bands. These values are in a good agreement with other tight-binding models \cite{Carr_Kaxiras_2018} as well as with the density functional theory calculation \cite{Lucignano_Cantele_2019, Carr_Kaxiras_2020, Uchida_Oshiyama_2014, Carr_Kaxiras_2018}. Achieving such results became possible only after incorporating the out-of-plane relaxation effects, which led to the enlargement of the band gaps, as well as widening of the flat band compared to the ideally flat MATBG. Notably, our model inherently lacks electron-hole symmetry. This feature contrasts with the continuum, Bistritzer-MacDonald model, where electron-hole asymmetry is obtained by adding momentum dependent inter-layer scattering terms \cite{Fang_Kaxiras_2019}. 

In Fig \ref{fig1}, we demonstrate the influence of the hBN substrate on the MATBG dispersion. In Fig. \ref{fig1} (c)  the black circles represent a pristine MATBG sample, while the red dots denote  the system on the substrate. One can notice that the presence of hBN has a negligible effect on the higher energy spectrum (i.e. remote bands) and the band gaps between the flat band and the remote bands remain unchanged. The impact of hBN becomes more apparent if we focus on the flat band, see Fig. \ref{fig1}(d). In this context, the presence of hBN leads to the opening of the gap at K-points and splitting of the degeneracies. 

\section{Quantum size effects in electronic structure of twisted bilayer graphene}
\label{s4}

  The electronic structure of the MATBG nanoribbons on an hBN substrate, considering various system widths, is analyzed in Fig. \ref{fig2}, where we plot the energy sub-bands as a function of the one dimensional momentum $k_{\rm 1D}$. In Fig. \ref{fig2} (a-c) bands in $200$ meV energy window are shown, while Fig. \ref{fig2} (d-f) provides a close-up view of the flat band. The color scale corresponds to the localization of the wave function. States localized on either of the sample's edges are represented in red, while bulk states that lie closer to the center of the system are displayed in blue. Given the symmetry of the spectrum with respect to $k_{\rm 1D} = 0$, only half of it is depicted. Dashed lines denote the band edges inferred from the band structure of the 2D infinite system. In the case of the smallest system consisting of 5 moir\'e unit cells, as depicted in Fig. \ref{fig2} (a), (d), the bulk states are predominant and  no edge states are apparent due to their strong hybridization with the bulk states. As the system size increases, edge states become more evident, clearly visible as red lines in Fig. \ref{fig2} (b), (e). This feature becomes even more pronounced for wider ribbons with band structures shown in Fig. \ref{fig2} (c), (f). We note that the wave function localization observed for the case of 20 moir\'e unit cells is in good agreement with the results for 85 moir\'e unit cells. Given the constant number of the edge states, the bulk band edges together with the gap widths, we have opted to  use the system consisting of 20 moir\'e unit cells in the subsequent calculations. We note that in Appendix A we confirmed the applicabilty of our method to study the edge effects by reproducing moir\'e flat-band breakdown shown recently in experiment in Ref. \cite{Yin_Qin_2022}

We also note two interesting features apparent at this stage of analysis. Firstly, there exists a distinct band gap around $E=0$ meV in all three cases, arising due to the presence of an hBN substrate.  In the gap region, no edge states link the bulk states above and below, resulting in a trivial gap with a Chern number $C=0$ \cite{Khalifa_Kaul_2023}. Secondly, we examined the flat sub-band degeneracy, which is constant for every wave vector k in the energy window between the remote bands. The number of states is always equal to the product of the moir\'e nanoribbon unit cell number and 8 (representing 2 bands, 2 valleys, 2 spins). This way there are always 20 states for each $k$ and each spin in Fig. \ref{fig2}(a),  80 in Fig. \ref{fig2}(e) and 340 in Fig. \ref{fig2}(f). This total number of states encompasses both edge and bulk states, however, the number of edge states remains constant regardless of the ribbon width. This observation implies that the two trivial edge states running below and the four states running above the flat band must originate from the flat band states. 

\subsection{Moir\'e-Hofstadter spectrum}
\label{s5}

\begin{figure*}\
\includegraphics[scale=0.4]{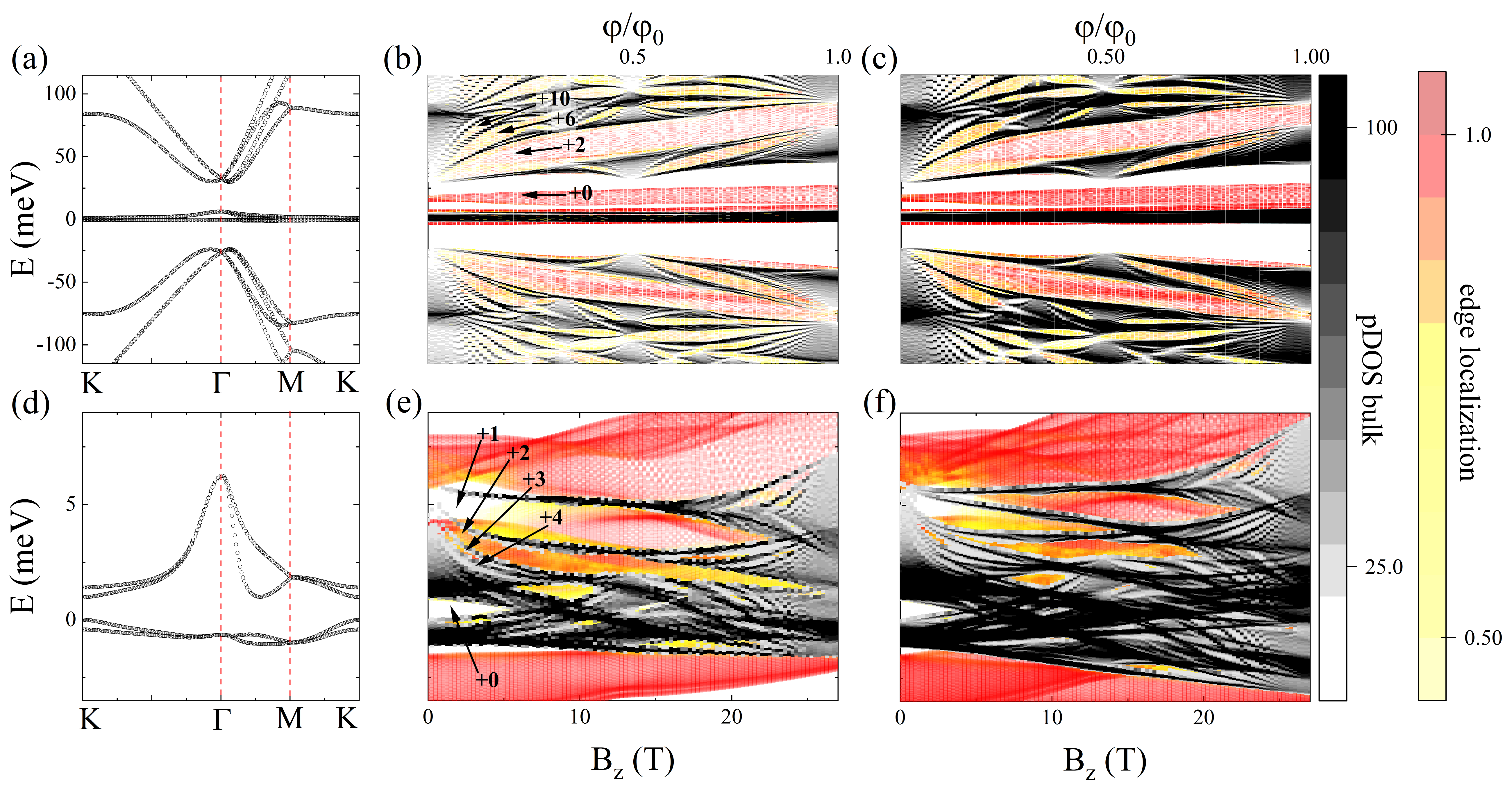}\
\caption{The band structure and Hofstadter spectrum of a MATBG ribbon on an hBN substrate. (a) Flat and remote bands within an approximately 200 meV energy window around the Fermi level along the $K-\Gamma-M-K$ line on the moir\'e Brillouin zone. (b) The Hofstadter spectrum for magnetic flux $\varphi / \varphi_{0}=[0,1]$, corresponding to magnetic field $B_{\rm z}\approx[0,27] $ T. This data is obtained from the nanoribbon density of states. The data points shown in greyscale represent projected density of states of the bulk, while the edge states are visualized in color. (c) Similar Hofstadter spectrum with Zeeman splitting included. The zoom-in to the energy window $E=[-5,10]$ meV of the flat band is shown in (d-f). Arrows highlight the gaps that we characterize by the in-gap Chern numbers.}\
\label{fig3}
\end{figure*}\
We analyze the influence of the perpendicular magnetic field on MATBG on top of hBN, exploring several aspects of the fractal spectrum. We consider magnetic fields in the range $\varphi/\varphi_{0}=[0, 1]$, corresponding to approximately $B_{\rm z}\approx [0, 27]$ T. The resulting Hofstadter spectrum is displayed in Fig. \ref{fig3} with a clear repeated pattern of energy gaps at different energy scales. Since our calculations are performed for a nanoribbon geometry, both bulk and edge states are present. The grayscale refers to the projected density of the bulk states, while the color scale indicates the degree of localization of the edge states. In Fig. \ref{fig3}(b) we show the fractal spectrum without the effect of the Zeeman splitting. First, we confirm that the width of the overall flat band remains constant and the flat bands do not mix with the remote bands up to one flux. From the remote bands, one can clearly observe the formation of the LLs for magnetic field $B_{\rm z} < 5$ T and their evolution as a function of $B_{\rm z}$. Flat energy band along M-K path around 80 meV and -80 meV without a magnetic field translates into the van Hove singularities that seem to be unaffected by small magnetic fields $B_{\rm z} < 5$T. There are no crossings of the edge state between the flat and remote bands, which leads to the conclusion that both of these gaps are trivial, and the in-gap Chern number is $C=0$. Analyzing the edge states crossing between the other LLs, we determine the in-gap Chern numbers for the other cases. The results of the edge state counting are marked with arrows pointing to the specific gaps. In Fig. \ref{fig3}(c) we present an analogical spectrum including the effect of the Zeeman splitting $E=\pm\frac{1}{2}g\mu_B B_{\rm z}$. In this energetic scale the only visible effect is a small broadening of the LLs, since the Zeeman splitting energy is of the order $\sim 10^{-2}$ meV/T.

In Fig. \ref {fig3}(e) and (f) we present close-ups of the Hofstadter spectra around the Fermi level focusing on a flat band region, once again with and without the Zeeman splitting, respectively. We analyze the in-gap Chern numbers corresponding to the number of edge states crossing it, concluding that the gap opened by hBN has no edge states crossing through it, and is, therefore, trivial with $C=0$. The band gaps within the split conduction band exhibit a sequence of Chern numbers: $C=1$, followed by $C=2$, $C=3$ and $C=4$. Notably, the top of the conduction band near the $\Gamma$ point has a negative effective mass, leading to the highest flat band LL decreasing in energy for an increasing magnetic field. This tendency deepens when we incorporate the Zeeman splitting, as seen in Fig. \ref{fig3}(f). In the case of the flat band, the Zeeman term corrections lead to the broadening of the energy bands related to the splitting of the Hofstadter spectrum.

\subsection{Analysis of Wannier diagrams}
\label{s6}
\begin{figure*}
\includegraphics[scale=0.37]{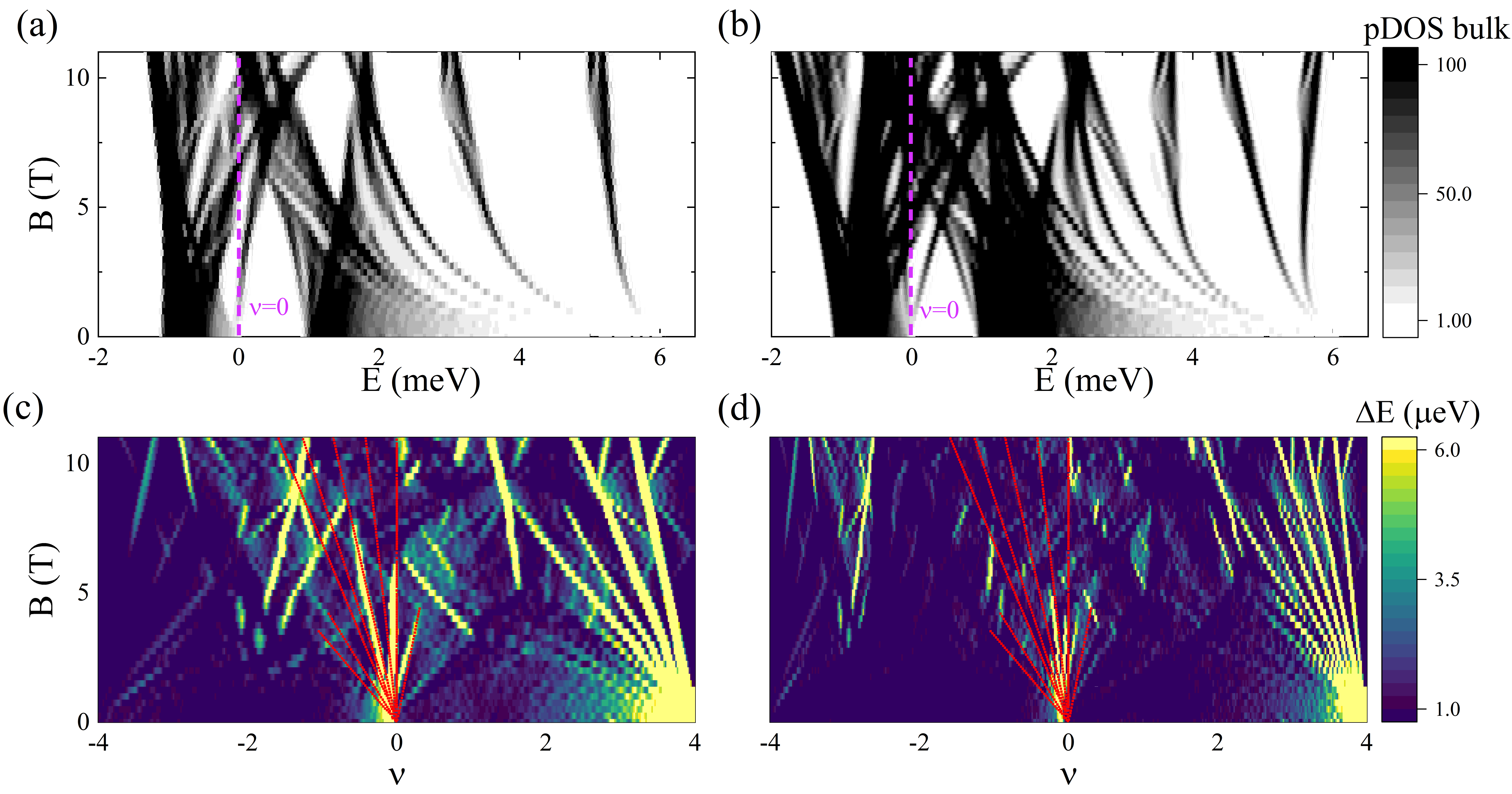}\
\caption{Wannier diagram of the hBN-MATBG ribbon flat band in the presence of a magnetic field. (a) Hofstadter spectrum for the bulk states of the flat band, extracted from data presented in Fig. 1 (d). (b) Hofstadter spectrum including Zeeman splitting. (c) Wannier diagram corresponding to the spectrum in panel (a). The x-axis measures the filling $\nu$ relative to the Fermi level. Filling -4 (4) indicates a completely empty (filled) flat band. The color scale is used to denote the energetic width of the band gaps $\Delta E$. Note that all gaps larger than 6 $\mu$eV are denoted by the same color. (d) Like (c), but this Wannier diagram is derived from the Hofstadter spectrum with Zeeman splitting included. The red dotted lines on (c) and (d) mark some of the gaps starting at $\nu=0$ extracted from the experimental data \cite{Pierce_Yacoby_2021}. All panels (a)-(d) have the same y-axis denoting the magnetic field $B_{\rm z}=[0,11] $ T.}\
\label{fig4}
\end{figure*}\
Through the analysis of the Hofstadter spectrum, a Wannier diagram can be obtained, as described in Section II C. For clarity, we first repeat our Hofstadter spectra, choosing a smaller magnetic field range in Fig. \ref{fig4}(a-b). Note that now the magnetic field $B_{\rm z}$ is plotted on the y-axis. We focus on the flat band and extract only the bulk projected density of states. In Fig. \ref{fig4}(a) and (b) we show the Hofstadter spectra without and with the Zeeman term, respectively.
 
The width of the energy gaps within the flat band is depicted in color in Fig. \ref{fig4}(c-d) as a function of the filling $\nu$ and the magnetic field $B_{\rm z}$. The red dotted lines mark some of the gaps that can be observed in the compressibility experiments (\cite{Pierce_Yacoby_2021}). Comparing this single particle result with the experimental data provides a direct means of identifying non-interacting, insulating states. We have achieved a good agreement with the experiment for fillings around $\nu=0$ and low magnetic fields. This suggests qualitatively correct mean-field, \textit{ab initio} based tight-binding picture around charge neutrality; correlations do not play an important role here.  On the downside, the vertical line at $\nu=0$, representing the largest electronic gap in the flat band spectrum, vanishes around $B_{\rm z}=5$ T, whereas in the experiments this gap remains open over the entire 0-11 T magnetic field range, see also Fig. 2 in Ref. \cite{Pierce_Yacoby_2021} for comparison. When the Zeeman splitting is included in our calculations, as shown in Fig. \ref{fig4}(d), some of the gaps have shifted to odd integer fillings, leading to a better agreement with the experimental results, e.g., the gaps visible in Fig. \ref{fig4}(c) around $\nu = -2$ for $B_{\rm z} > 3 $ T have shifted to $\nu = -3$ also for $B_{\rm z} > 3 $ T. However, since the number of gaps has doubled and their widths have decreased, some of the previously prominent features have become less distinct, e.g., note the rightmost gap beginning at $\nu = 0$ and closing around $B_{\rm z} = 5$ T. It is well visible in Fig. \ref{fig4}(c) in agreement with the experimental data, but becomes much less clear after including the Zeeman splitting, see Fig. \ref{fig4}(d). Hence, studying both scenarios - with and without the Zeeman splitting - can provide valuable insights and enable us to better determine which gaps are not related to the effects of interaction. While we leave the electrostatic and many-body interactions for a future work, we anticipate a widening of the gaps around integer fillings. Moreover, there is a very strong feature below $\nu=4$ stemming from the well-separated top flat band LL, which is not observed in the experiments. This suggests a significant renormalization beyond the current understanding of the energy dispersion around the $\Gamma$ point in the magnetic field due to the presence of interactions or magnetic fields.

\subsection{Conventional, mixed and moir\'e state properties}
\label{s7}

 Within an energy window of approximately 100 meV around charge neutrality, two types of bulk magnetic bands with highly suppressed kinetic energy emerge. The first are the moir\'e flat bands, which in the real space picture correspond to states mostly localized in the AA stacked regions \cite{Bistritzer_MacDonald_2011a} of the twisted bilayers. The second feature more conventional Landau levels, which generally extend across the entire moir\'e unit cell and are not restricted to the 'moir\'e centers'. It is currently unclear how these two types can be modeled simultaneously in low energy theories and how they are mutually influenced. For example, several works show different behaviours of the Hofstadter butterflies and related different Chern numbers of the gaps \cite{Bistritzer_MacDonald_2011, Moon_Koshino_2012, Lian_Bernevig_2020, Guan_Kruchkov_2022}. Even less is known about the microscopic properties of wavefunctions of these levels, especially about chiral states connecting bulk bands, localized on the edges of TBG samples. Since this knowledge is crucial for understanding the topological aspects of MATBG, the interacting problem, Chern insulator states, potential re-entrant superconductivity, and quantum geometric properties, \cite{Torma_Bernevig_2022, Herzog-Arbeitman_Bernevig_2022a, Herzog-Arbeitman_Bernevig_2022b, Herzog-Arbeitman_Bernevig_2022c, Guan_Kruchkov_2022, Shaffer_Santos_2022} in our work we provide general microscopic features of both bulk and edge states of MATBG in a magnetic field. 
 
It is important to experimentally probe the edge state physics, e.g. extend scanning tunneling spectroscopy studies of the breakdown of the moir\'e flat bands \cite{Yin_Qin_2022} to the identification of the Chern-moir\'e edge states, and extend transport studies of possible zero field Chern phase \cite{Serlin_Young_2020} to non-local Hall measurements and to a finite magnetic field. 
The local probing of competing ground state wave functions is still in its early stages for both experimental \cite{Nuckolls_Yazdani_2023} and theoretical investigations \cite{Calugaru_Bernevig_2022, Hong_Zaletel_2022}.

Other exciting avenues for probing edge currents are superconducting quantum interferometry \cite{Fortin-Deschenes_Xia_2022} and electron spin resonance edge probes \cite{Sichau_Blick_2019, Singh_Blick_2020, Morissette_Li_2023, Tiemann_Blick_2022}. Although some works have discussed the moir\'e edge states \cite{Moon_Koshino_2012, Fleischmann_Shallcross_2018, Liu_Dai_2019b, Fujimoto_Koshino_2021, Khalifa_Kaul_2023}, they primarily centered on the effective models or angles greater than the magic angle. Studies using large-scale atomistic tight-binding models \cite{TramblydeLaissardiere_Magaud_2010, SuarezMorell_Barticevic_2010, TramblydeLaissardiere_Magaud_2012, Sboychakov_Nori_2015, Fang_Kaxiras_2016, Kang_Vafek_2018, Rademaker_Mellado_2019, Lothman_Black-Schaffer_2022, Kang_Vafek_2023, Kang_Vafek_2023b, Miao_Dai_2023} have predominantly focused on the electronic structures, relaxation effects and interactions, serving as a basis for the construction of low-energy models. 
\begin{figure}\
\includegraphics[scale=0.22]{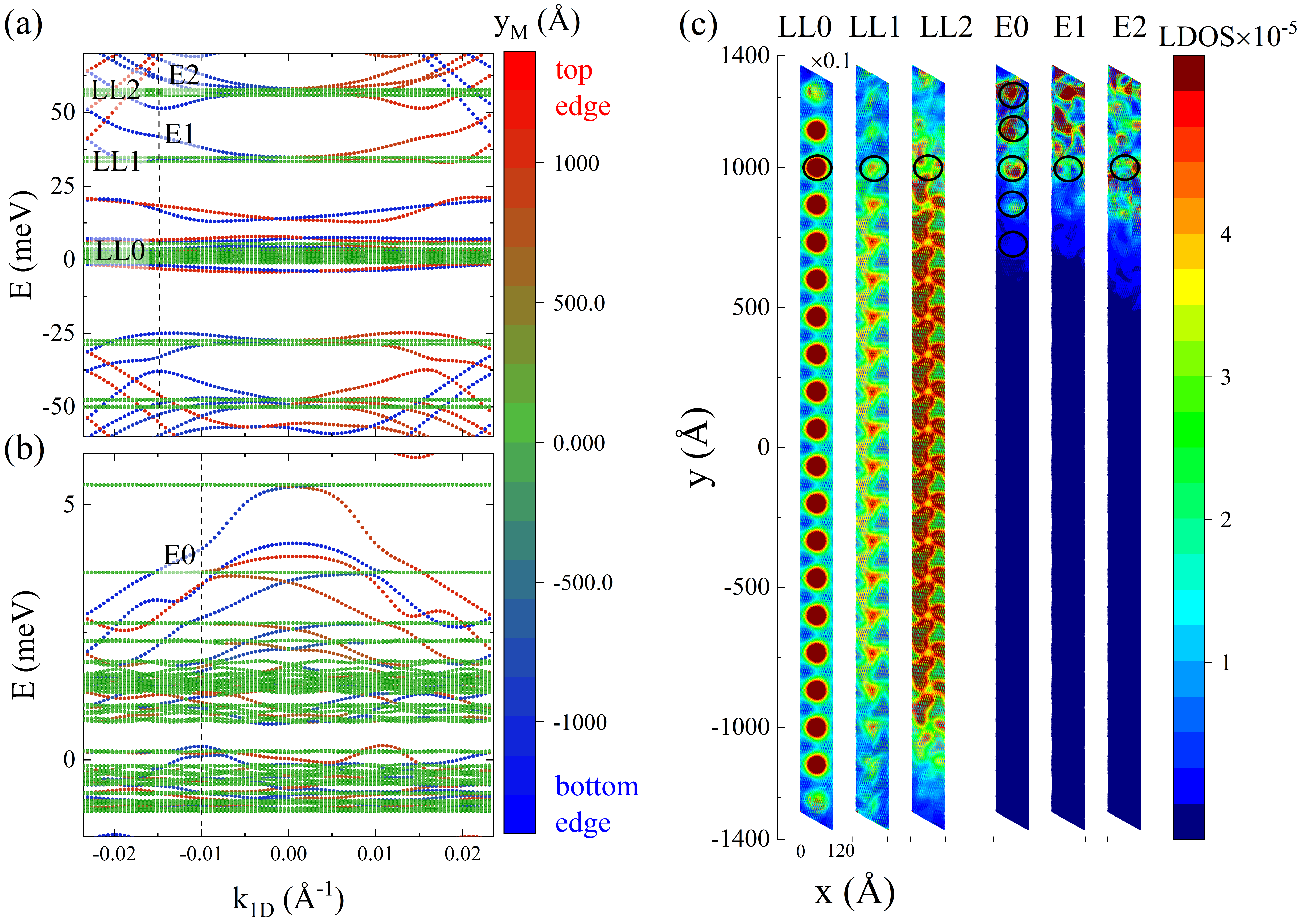}\
\caption{Electronic structure and wave functions of the hBN-MATBG ribbon in the magnetic field. (a) Band structure of the 20 moir\'e unit cells wide hBN-MATBG nanoribbon in perpendicular magnetic field $B_{\rm z} = 4$ T without Zeeman splitting. Both LLs (LL0, LL1, LL2) and edge states (E1, E2) are shown. The color scale denotes the wave function localization, with top (bottom) edge states marked by red (blue) dots. Green dots mark bulk states predominantly localized in the center of the sample. (b) Zoom-in into the flat band, revealing its substructure in the magnetic field, again with the moir\'e edge states (E0) crossing between the moir\'e LLs. (c) LDOS for the bulk and edge states, which are marked on (a) and (b). The specific values of $k_{1D}$ for which they were calculated are indicated with a black dashed line. The left side of the plot (c) shows selected bulk states – the flat band (LL0), the first LL (LL1) and the second LL (LL2). On the right, the edge states are shown  - E0, which is localized within the flat band, E1 which connects LL1 and LL2, and E2, which connects LL2 with LL3 (the 3rd LL). The black circles correspond to the moir\'e centers, which are fixed around the AA stacked atoms in the unit cell. Note that the color scale for LL0 is scaled by 0.1 compared to the rest of the LDOS plots in (c).}\
\label{fig5}
\end{figure}\

Here, we investigate the microscopic properties of the MATBG wavefunctions in a magnetic field for the 20 moir\'e unit cells nanoribbon and a representative magnetic field $B_{\rm z}=4$ T. The 1D band structure is shown in Fig. \ref{fig5} (a-b). We consider two energy windows, with $\Delta E\approx 150$ meV in Fig. \ref{fig5} (a), followed by a zoomed-in view of the flat band region ($\Delta E \approx 10$ meV) in Fig. \ref{fig5} (b). In both cases, we classify the ribbon states according to their localization. States colored in green represent the ones localized within the center of the sample, while states colored in red and blue correspond to those localized on the top or bottom edges, respectively. The bulk states form flat bands, i.e. the ribbon LLs, between some of which the edge-localized states cross the gaps. This is a characteristic signature of non-zero Chern number. We note, that not all gaps are crossed by the states connecting bulk bands, suggesting the existence of trivial edge states. 

Before we delve into the details of the real-space properties of the ribbon states, let us precisely define the meanings of conventional-, moiré-, and mixed-type states, classifications we use in the subsequent paragraphs. We label states as 'conventional' if the LDOS is uniformly spread over the sample and resembles a typical LL distribution. 'Moiré' states are characterized by the LDOS primarily concentrated around the moiré centers. For 'mixed' states, LDOS is neither concentrated only around the moiré centers nor uniformly spread across the sample.

To analyze the real space properties of the ribbon states, we have chosen three bulk LLs - the flat band (LL0) around $E=0$ meV, the first LL (LL1) around $E=30$ meV, and the second LL (LL2) around $E=60$ meV. Our investigation of the microscopic properties of the electronic charge densities is summarized in Fig. \ref{fig5}(c), which shows the LDOS calculation and its spatial distribution. Within this analysis, we focus on three consecutive LLs, which represent three different types of bulk states.  The first stripe shows the flat band, LL0, which as expected is a "moir\'e" type state - the LDOS is concentrated predominantly around the region of the AA stacking in the real space. The second type, LL1, which is separated by a $\sim 30$ meV gap from the flat band, still strongly feels the LL0 influence, and also exhibits LDOS concentration around the moir\'e centers. This result is surprising, since LL1 is a regular LL and, in principle, should present a uniform spread of the wave function. The fact that it doesn't, suggests a stronger influence from the flat band than anticipated, potentially observable in a broader energy range. This is an example of a "mixed" state. Moving to LL2, the "conventional" bulk state, we observe a more uniform LDOS spread, as one would expect. However, there is also an opposite trend present - LDOS avoids the moir\'e centers, notice yellow dots in the center of moir\'e unit cells for LL2 in Fig. \ref{fig5}(c). For higher LLs, the pattern is similar to that of LL2, with an even more evenly distributed wave function.
A similar analysis can be conducted for the edge states. Once again, we identify three distinct types of edge states. Those states are indicated  in Fig. \ref{fig5}(a) and \ref{fig5}(b) by: E0 which resides within the flat band; E1 connecting LL1 with LL2; and E2 linking LL2 with higher-lying LL3. The right side of the Fig. \ref{fig5}(c) shows the localization of the wave function for these three cases, and in particular what is its relation to the moir\'e centers marked with black circles.  E0 is predominantly localized within the moir\'e centers close to the edge of the sample, and the trivial edge states in the energy gap between LL0 and LL1 around $E=25$ meV behave in a similar manner. In the case of E1, the AA stacking localization can still be spotted, although the effect is less prominent. As for E2, there is no correspondence between the wave function localization and the moir\'e centers position. This is to be expected, since this state is separated by over 50 meV from the flat band, and moir\'e potential does not play a role here. 


\subsection{Evolution of the microscopic wave function's properties}
\label{s8}
\begin{figure}\
\includegraphics[scale=0.3]{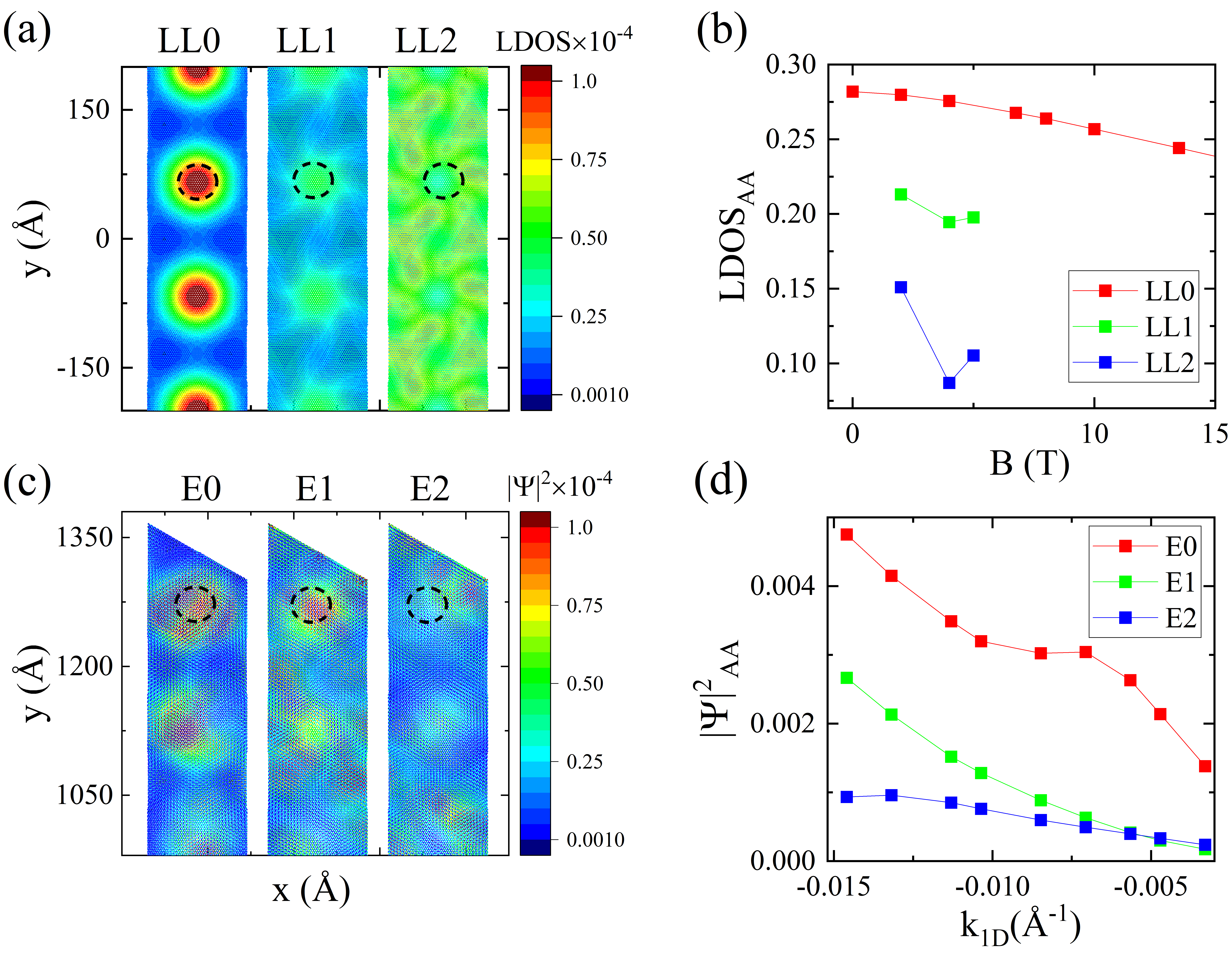}\
\caption{LDOS within moir\'e AA centers for bulk and edge states in a magnetic field for an hBN-MATBG ribbon. (a) Real-space distribution of LDOS for selected bulk LLs (LL0-LL2) shown around the center of the sample for $B_{\rm z} = 4$T. (b) Corresponding integrated LDOS calculated for the moir\'e center defined as the region within a black circle centered around AA stacked atoms with the radius $r= 25$ \AA. (c) Wave function density $|\Psi|^2$ plot of selected edge states, from left: E0, which lies within the flat band, E1 connecting LL1 and LL2, and E2, linking LL2 with LL3. (d) Similar to (b), $|\Psi|^2$ analysis, with integrated density calculated for the moir\'e center closest to the edge of the ribbon.}\
\label{fig6}
\end{figure}\
We now turn into the analysis of how the LDOS changes in response to the magnetic field. For concreteness, we define a moir\'e center as the region within the dashed circle marked in Fig. \ref{fig6}(a). This figure shows that there are strong differences between LL0, LL1, LL2, and those three states should be easily distinguished in scanning tunneling spectroscopy type experiments. We investigate how much of the wave function is localized within the moir\'e center, and how this distribution evolves with an increasing magnetic field. The outcomes of this calculation are presented in Fig. \ref{fig6}(b), where we quantitatively compare the wave function localization for LL0, LL1, and LL2. We also show, that in the flat band case, as the magnetic field increases, the wave function flows out of the moir\'e center and distributes more evenly over the sample, as the effect of the magnetic field becomes dominant over the effect of the moir\'e potential. Therefore, one can observe the decreasing character of the red line in Fig. \ref{fig6}(b). In the case of LL1 and LL2, denoted by green and blue lines respectively, there is a similar trend present, however, it cannot be observed for a larger range of $B_{\rm z}$, since with the increase of the magnetic field these LLs are not unequivocally defined.

We have also focused on the moir\'e center nearest to the edge of the sample, as depicted in Fig. \ref{fig6}(c). Here, we investigate how much of the wave function localizes in the immediate vicinity of this center as a function of the wave vector $k_{\rm 1D}$. These findings are summarized in Fig. \ref{fig6}(d), where a clear distinction between E0, E1 and E2 is shown. This observation reinforces our classification of the three types of edge states. The evolution of these edge states as a function of $k_{\rm 1D}$ looks nontrivial and warrants a more extensive study in the future. Their behaviour as a function of $k_{\rm 1D}$ is characterized by qualitative differences, with E0 exhibiting a local minimum, E1 decaying exponentially, while E2 remaining mostly flat. These features could potentially be verified by experiments. 

\section{Conclusions}

In this work, we analyzed the microscopic properties of magic angle twisted bilayer graphene on top of hBN under the influence of a magnetic field. We have studied systems consisting of up to 1 million atoms (85 moir\'e unit cells) and have established that for a ribbon's width larger than 20 unit cells, hybridization between the wavefunctions of the edge states is negligible. For such wide ribbons, we studied the Hofstadter spectrum and determined the in-gap Chern numbers from the number of chiral edge states crossing the energy gaps. At low magnetic field, a corresponding Wannier diagram has been obtained and used to identify noninteracting, insulating states, finding a qualitative agreement with the experiments. We have analyzed a competition between moir\'e localization potential and the effect of a magnetic field. The microscopic properties of the wave functions were determined and three types of the bulk and edge states were identified, namely moir\'e, mixed and conventional states. We have examined their evolution as a function of the magnetic field and the wave vector. While our analysis was primarily restricted to single-particle physics without electrostatic corrections for doping away from $\nu=0$, we anticipate that our findings can provide a valuable guidance for future scanning tunneling microscopy measurements and help to experimentally establish properties of microscopic wave functions in twisted bilayer graphene.  

\section*{Acknowledgments}

The authors thank Nicolás Morales Durán, Gaurav Chaudhary, Weronika Pasek and Tobias Wolf for useful discussions. AWR, DM and PH were supported by NSERC Discovery Grant No. RGPIN 2019-05714, the QSP-078 and AQC-004 projects of the Quantum Sensors and Applied Quantum Computing Programs at the National Research Council of Canada, University of Ottawa Research Chair in Quantum Theory of Materials, Nanostructures, and Devices, and Digital Research Alliance Canada with computing resources \cite{alliance_canada}.  
MB and RT acknowledge support by the Deutsche Forschungsgemeinschaft (DFG, German Research Foundation) through the Würzburg- Dresden Cluster of Excellence on Complexity and Topology in Quantum Matter – ct.qmat Project-ID 390858490-EXC 2147.
PP acknowledges support from the
Polish National Science Centre based on Decision No. 2021/41/B/ST3/03322. 

\section*{Appendix A: Moir\'e flatband breakdown}

\begin{figure}\
\includegraphics[scale=0.42]{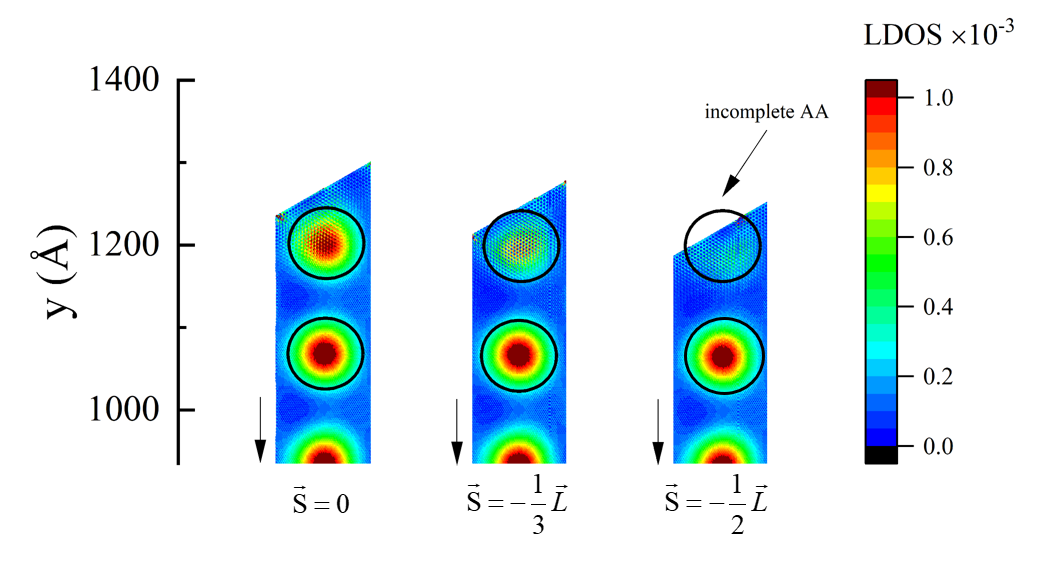}\
\caption{LDOS near the top edge of 20 moiré unit cell ribbons for different choices of boundary termination. Different cuts are parametrized by the shift vector S. Black circles denote the AA regions.}\
\label{fig10a}
\end{figure}\

\begin{figure}\
\includegraphics[scale=0.38]{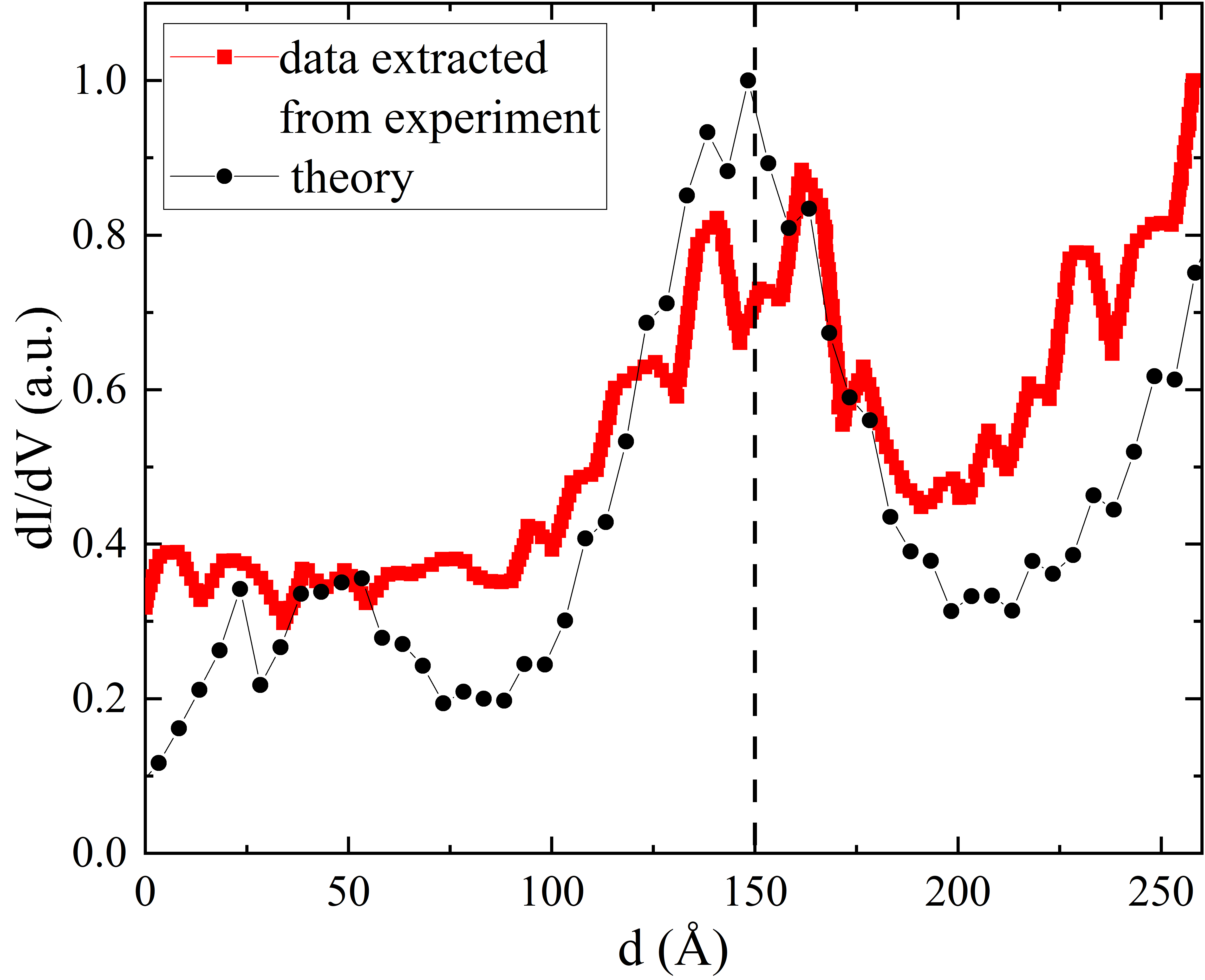}\
\caption{ Comparison of $dI/dV$ spectrum near the edge of the MATBG sample extracted from Ref. \cite{Yin_Qin_2022} with our spatially integrated LDOS calculation shows the effect of flat-band breakdown. Distance $d$ is measured from the edge of the sample. The dashed line indicates the center of one of the AA regions.}\
\label{fig10b}
\end{figure}

Recent experiment \cite{Yin_Qin_2022} has shown that the characteristic flat band LDOS localization around the moir\'e centers of the superlattice remains intact even close to the edge, as long as the sample contains a complete moir\'e spot. However, if the edge of the sample cuts through the AA region, the flat band LDOS will no longer be localized in that area. Our model correctly captures these features. In Fig. \ref{fig10a}, we present three different ways of terminating our nanoribbon. In the first case, we preserve the whole moir\'e center at the edge ($\vec{S}=0$), while in the second and third, we shift our original unit cell by a vector $\vec{S} = -1/3\vec{L}$ and $\vec{S} = -1/2\vec{L}$, respectively ($\vec{L} = \vec{L}_1 + \vec{L}_2$, see Eq. 1). One can notice that in the first case, there is a complete moir\'e center localized close to the edge, but for the other two choices of the edge, the LDOS fades away in that region. We have compared our results with the experimental data extracted from Ref. \cite{Yin_Qin_2022} in Fig. \ref{fig10b}. Both in the experimental result and our spatially integrated LDOS result, there is a clear change in the electron density when the edge cuts through the AA moir\'e region. This confirms that our model is capable of capturing local physics near the edges of MATBG system under study.
\
\section*{Appendix B: Effect of hBN substrate on Hofstadter spectrum and Wannier diagrams}
\begin{figure*}\
\includegraphics[scale=0.38]{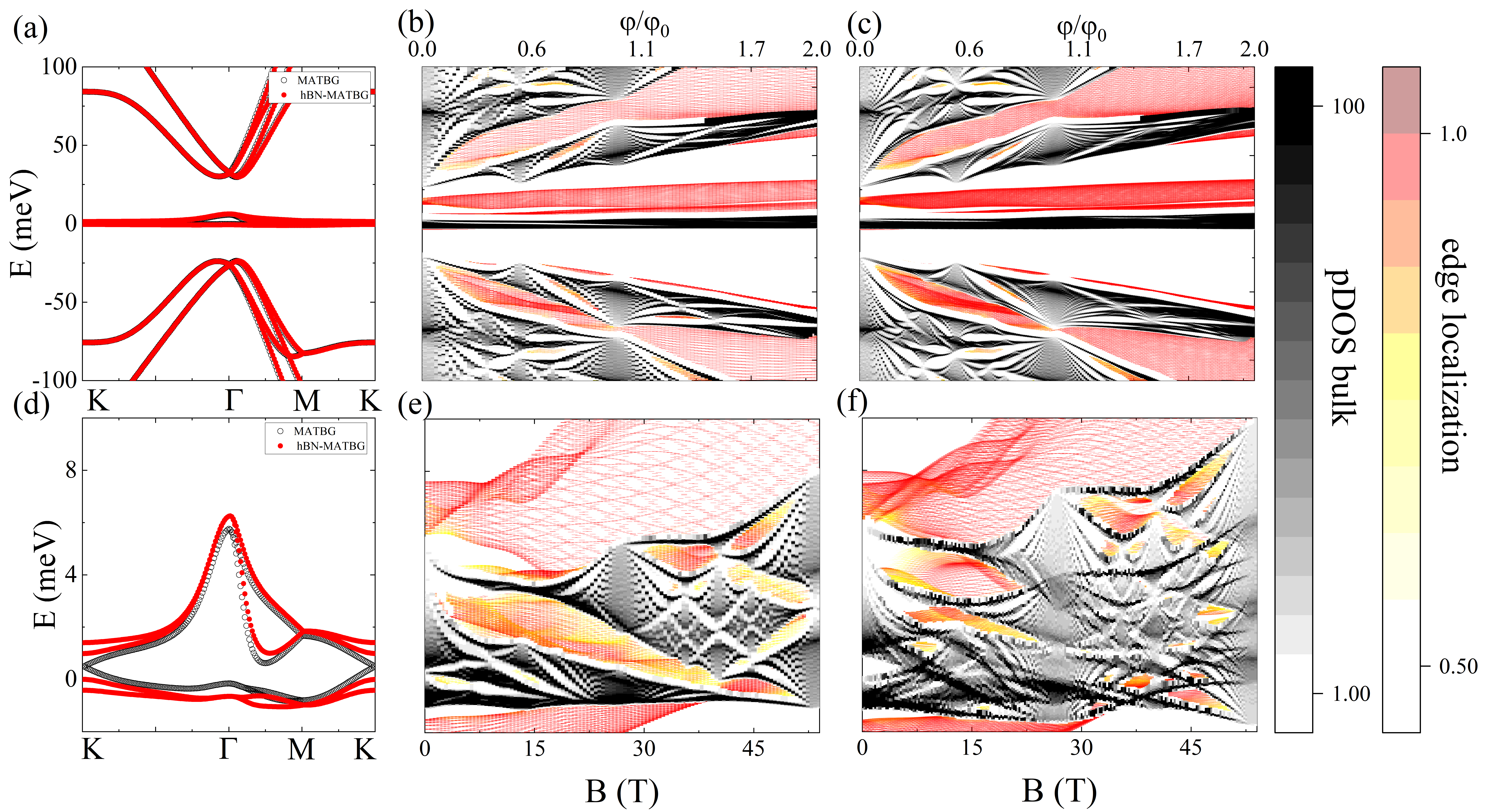}\
\caption{The influence of the hBN substrate on the band structure and the Hofstadter spectrum. (a) Flat and remote bands around the Fermi level. The black circles represent a pure MATBG sample, while the red dots depict the impact of the  hBN substrate. The Hofstadter spectrum (b) without  and (c) with the hBN substrate effect.  (d-f) Zoom-in to the flat band. Panels (e) and (f) provide a corresponding close-up view of the MATBG fractal spectrum without and with the effect of hBN, respectively. }\
\label{fig7}
\end{figure*}\
Now we elaborate further on the role of an hBN substrate. As outlined in Section II, we consider the presence of hBN on only one side of the sample and model its influence by introducing a staggered potential. It is important to emphasize that if one were to add microscopically realistic layer of hBN, there would be a number of additional effects. The inherent lattice mismatch between hBN and graphene typically gives rise to an extra moir\'e pattern with a period comparable to that of MATBG, which could potentially affect the flat bands. One of the current experimental approaches is to twist the hBN layer with respect to MATBG by a small angle to match their moir\'e patterns in order to maximally flatten the bands. Consequently, our treatment of hBN should be treated as a simplified one. To gain a clearer understanding of its impact, we here isolate the influence of hBN on the Hofstadter spectrum. In Fig. \ref{fig7} (a) and (d) we reproduce the bulk states’ band structure in $B_z=0$ T aiding in identifying the sources of high densities of states in the Hofstadter spectrum for $B_z \approx 0 $ T. Comparison of the fractal spectra in a large energy window shows a similar trend, demonstrating the same features for a sample Fig. \ref{fig7} (b) without and (c) with the hBN substrate. However, a closer examination of the zoomed-in spectrum in Fig. \ref{fig7} (e-f) reveals significant differences between (e) MATBG  and (f) hBN-MATBG. Firstly, there is a trivial gap with no edge states crossing through it, which emerges around the zero energy for magnetic fields between 0-5 T. The spectrum displays numerous additional gaps in general, resulting in a much richer substructure of the flat band compared to pristine MATBG. These gaps are generally narrower and remain open within limited ranges of the magnetic field values.

\begin{figure*}\
\includegraphics[scale=0.37]{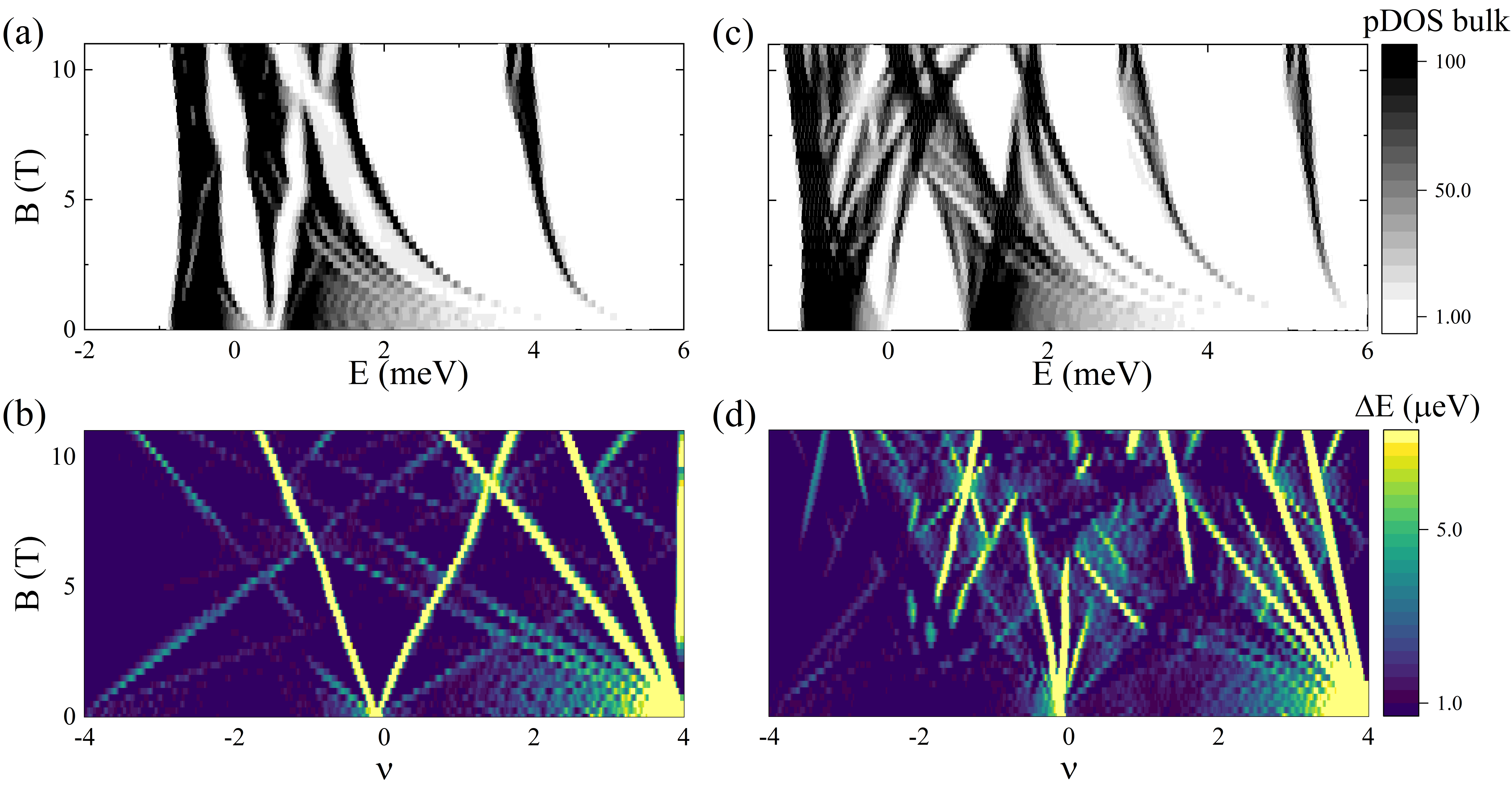}\
\caption{The role of an hBN substrate on the Wannier diagram in the flat band of MATBG. (a) Hofstadter spectrum illustrating the bulk states of the flat band in MATBG without the hBN substrate. (b) Corresponding Wannier diagram for the spectrum shown in  (a). Characteristic V-shape picture around $\nu=0$ is seen in the experiments \cite{Das_Efetov_2021,Yu_Feldman_2022}. (c) Fractal spectrum, similar to (a), but accounting for the effect of hBN. (d) Wannier diagram corresponding to (c), using the same color scale as (b). }\
\label{fig8}
\end{figure*}

The effects associated with the influence of hBN on MATBG become even more evident when examining the Wannier diagrams. Panels (a) and (b) in Fig. \ref{fig8} pertain to a pristine MATBG sample, while panels (c) and (d) correspond to MATBG on hBN. Once again, it is apparent that the presence of hBN induces the emergence of numerous additional bandgaps, particularly for larger magnetic fields. The Wannier diagram depicted in Fig. \ref{fig8} (b) can be compared to the compressibility measurements for unaligned devices \cite{Das_Efetov_2021,Yu_Feldman_2022}, wherein a similar V-shape stemming from the charge neutrality point can be observed.

\
%
%
\bibliographystyle{apsrev4-2}
\bibliography{ver1_TBG_14May2023}
\end{document}